# Highly tunable charge transport in defective graphene nanoribbons under external local forces and constraints: A hybrid computational study

**Mahnoosh Rostami[a], Isa Ahmadi[a], Farhad Khoeini[b*],**

[a]Department of Mechanical Engineering, University of Zanjan, 45195-313, Zanjan, Iran
[b]Department of Physics, University of Zanjan, 45195-313, Zanjan, Iran

**Abstract**

In this paper, we propose a combined modeling of molecular mechanics (MM) and the tight-binding (TB) approach, which enables us to study the effect of factors such as external local forces, constraints, and vacancy defects on electronic transport properties of nanomaterials. Nanostructures selected in this work are armchair graphene nanoribbons (AGNRs). According to this method, the nanostructure is modeled as a frame, and the beam element is applied for illustrating the covalent interatomic interactions in bonds. In our calculations, the terms of torsional, stretching, and bending energies are considered. The selected pristine nanoribbon is a metal, and the purpose of this study is to find the effects of mechanical loading, the vacancy defects and their positions on the electrical conductance of the structure. We observe that the presence of vacancy defects in the structure leads to the opening of an energy gap, which changes the phase of the nanostructure from metal to the semiconductor. We find that with increasing the number of point defects, the energy gap size of the strained system grows. Besides, increasing the magnitude of the local force reduces the conductance, and the energy gap of the system. By changing parameters such as the number of point defects and magnitude local forces, the transport gap of the system can be controlled. The results of this research may be useful in the design of nanoelectromechanical systems.

## 1. Introduction

Graphene is a two-dimensional (2D) lattice of carbon atoms with zero-bandgap [1-6]. After synthesizing graphene as the first known 2D material, a large class of 2D materials and their ribbons such as phosphorene [7-11] boron structures [12-14], silicene, stanene, and graphyne have been studied [15-20].

Carbon nanostructures have superior properties such as high inherent mobility [21], high electrical and thermal conductivity [3, 22-24], high stiffness and strength [25, 26], which makes them appropriate materials for membrane technology, optics, electronics, and sensor applications [15, 27, 28].

In theoretical studies of mechanical analyzing of the nanostructures, the following methods can be employed: quantum mechanics, molecular mechanics, molecular dynamics (MD), and continuum mechanics (CM), which among them, the MD and the MM simulations are suitable methods for studying the thermal and mechanical behavior of a nanostructure [3, 29].

With the help of these methods, many studies have been done on the thermal, mechanical, and electrical properties of graphene structures. Controlling the properties of these structures is possible by applying parameters such as defects or tensile strain [27, 30]. In this context, Lehmann et al. [31] investigated the role of uniaxial strain and defects on the electronic transport properties of graphene nanoribbon. They found that an energy gap is opened by applying the strain, and its size grows by increasing the concentration of vacancy defects. Recently, Yousefi et al. [32] investigated the thermal properties of the graphene sheet by introducing the nanoporous on it. They observed the presence of nanoporous reduces the thermal conductivity. Similarly, Nazarloo et al. [33] used the nonlinear modified Morse model to investigate the mechanical behavior of the graphene sheet. Nonlinear spring was applied for analyzing the finite element model (FEM) and obtaining the elastic moduli of structure. They found the effects of size and aspect ratio of structure on elastic properties. Recently, Mohammadi et al. [34] studied Young's modulus of thin graphene sheets under uniaxial stress by the MD method.

As we know, graphene has limited electronic applications due to its zero bandgap [35]. One of the strategies for changing the electronic properties of graphene is to engineer the transport gap size of the structure [3, 12]. Li et al. [36] investigated the effects of uniaxial and shear strains on the conductance of graphene and graphene nanoribbons. The applied approach was a combination of DFT and TB methods. Their results illustrate that no bandgap is generated in the strained graphene. They have shown that the energy gap of AGNRs is insensitive to shear strain, while it is varied by the uniaxial strain in a zigzag ribbon.

Recently, Shakeri et al. [37] studied the role of stone-wales defects in AGNRs and their corresponding transmission characteristics. They applied the TB model for simulating the structure and four distribution functions to represent the role of the number and position of defects. They found that a bandgap is opened, and its width grows with increasing the density of defects.

Many studies have been done on the nonlocal strain, while the local strain has been less investigated. Pereira et al. [38] explored the effect of local strain on the electronic structure of graphene by the tight-binding method. They illustrated that local strain could be easily designed to generate electron beam collimation, quantum wires, and confined states in graphene. Gui et al. [39] investigated the electronic properties of graphene under local strain with DFT simulation, and observed that appropriate local strain could tune the bandgap. The results of these two studies are consistent with each other. Thus, the local strain is a simple way to open a bandgap in graphene for practical electronic applications.

**To the best of our knowledge, our study is the first report for considering the effect of local force on electronic transport properties of graphene nanoribbons.** This study investigates the electrical conductance of graphene nanoribbons in the presence of local mechanical loading and vacancy defects by the molecular structural mechanics (MSM) approach, which is based on the MM method. We use the TB method to study electrical properties.

The organization of the paper is as follows. In section 2, we propose a hybrid model of the MM and TB to study the effect of factors such as external local forces and vacancy defects on the electronic transport properties of a nanosystem. In section 3, our results are presented in two cases of pure and defective systems. Finally, we conclude with a summary in section 4.

## 2. Model and method

In this section, we will first do mechanical modeling and then electrical one. The studied system is a graphene-based armchair ribbon connected to two electrodes.

### 2.1. Mechanical modeling

The structural molecular mechanics approach, at the macroscopic level, establishes the connection between molecular mechanics and classical mechanics methods. In this paper, an armchair graphene nanoribbon is considered as a frame-like structure, and the atomistic finite element method (AFEM) is applied for modeling the structure. The covalent bonds between carbon atoms are represented by beam elements, and atoms are treated as nodes. The physical parameters of the beam elements have already been derived from the relation between the potential energy of graphene and the strain energy of the nanostructure.

The total potential energy for a nanoribbon is given by [40-43]

$$U = \sum U_r + \sum U_\theta + \sum U_\emptyset + \sum U_\omega + \sum U_{vdv} + \sum U_{EL}, \qquad (1)$$

where, $U_\emptyset$, $U_\theta$, $U_r$ represent the terms of the energy-related to torsion, bending, and stretching in bonds, respectively. The interatomic interaction from the molecular mechanics point of view is shown in Fig. 1(b). $U_\omega$ demonstrates the energy corresponding to out-of-plane torsion, $U_{EL}$ and $U_{vdv}$ represent the electrostatic and van der Waals energies, respectively. The electrostatic interaction is omitted due to the large difference between the mass of electron and nucleus in the carbon atom. The van der Waals energy is also ignored because of its non-bonded nature. Thus, at the atomic scale, the harmonic approximation of potential energy is appropriate for describing the bond interactions [44, 45].

In the range of small deformation, the total potential energy for the beam can be defined as [40, 46]

$$U_r = \frac{1}{2}K_r(r - r_0)^2 = \frac{1}{2}K_r(\Delta r)^2, \qquad (2\text{-a})$$

$$U_\theta = \frac{1}{2}K_\theta(\theta - \theta_0)^2 = \frac{1}{2}K_\theta(\Delta\theta)^2, \qquad (2\text{-b})$$

$$U_\tau = U_\emptyset + U_\omega = \frac{1}{2}K_\tau(\Delta\emptyset)^2. \qquad (2\text{-c})$$

$K_r$, $K_\theta$, and $K_\tau$ are the force constants for the stretching, bending, and twisting the covalent bonds, respectively. $\Delta r, \Delta\theta$, and $\Delta\emptyset$ are the increase of bond length, bond angle change, and deviation of the dihedral angle from the equilibrium state. The cross-section of the beam element is circular. Thus, the moment of inertia is $I_x = I_y = I$ and the force field constants are considered as $K_r = 6.52 \times 10^{-7}$ N/nm, $K_\theta = 8.76 \times 10^{-10}$ N.nm/rad$^2$, $K_\tau = 2.78 \times 10^{-10}$ N.nm/rad$^2$, respectively. The structural and mechanical required parameters for modeling the beam element, like Young's modulus and Shear modulus, can be obtained by equalizing the molecular potential with the total strain energy of the structure. The essential data to calculate the beam stiffness matrix are presented in Table 1. The strain energy for the beam element, in the presence of an axial force, F, is given by [47-49]

$$U_N = \frac{1}{2}\int_0^L \frac{P^2}{EA}dL = \frac{1}{2}\frac{P^2 L}{EA} = \frac{1}{2}\frac{EA}{L}(\Delta L)^2, \qquad (3)$$

where $U_N$, E, are stretching energy and Young's modulus of the beam, respectively. Besides, A, L, and ΔL are the area of the cross-section, the element length, and the change of bond length. Likewise, the strain energy in pure bending and torsion states can be calculated by the following equations:

$$U_M = \frac{1}{2}\int_0^L \frac{M^2}{EI}dL = \frac{2EI}{L}\alpha^2 = \frac{1}{2}\frac{EI}{L}(2\alpha)^2, \tag{4}$$

$$U_T = \frac{1}{2}\int_0^L \frac{T^2}{GJ}dL = \frac{1}{2}\frac{T^2 L}{GJ} = \frac{1}{2}\frac{GJ}{L}(\Delta\beta)^2. \tag{5}$$

where I and α are the moment of inertia and the angle of rotation at the end of the beam, respectively. Besides, G, J, and Δβ are the shear modulus of the beam, the polar moment of inertia, and the relative rotation at the end of the element, respectively. After calculating the total strain energy for the beam element, the structural and mechanical parameters associated with the force constants can be written as [47, 50, 51]:

$$K_r = \frac{EA}{L}, \tag{6-a}$$

$$K_\theta = \frac{EI}{L}, \tag{6-b}$$

$$K_\tau = \frac{GJ}{L}. \tag{6-c}$$

Table 1. Required Parameters for Calculating beam stiffness matrix

| Force field constants | Structural, mechanical parameters | |
|---|---|---|
| $K_r = 6.52 \times 10^{-7}$ N/nm | E=5.49 TPa | A=0.0678 (nm$^2$) |
| $K_\theta = 8.76 \times 10^{-10}$ N.nm/rad$^2$ | G=8.74 TPa | L=0.142 nm |
| $K_\tau = 2.78 \times 10^{-10}$ N.nm/rad$^2$ | J=2I=2.29e$^{-5}$ (nm$^4$) | d=0.147 nm |

Fig. 1(a) shows a schematic of an armchair graphene nanoribbon and its boundary conditions, which are described in detail in Section 2. The ribbon size is a=7.53 nm, and b=0.86 nm with clamped boundary conditions at sides x=0 nm and x=7.53 nm (shown in green area and sites) and other free sides. The position of carbon atoms relative to each other (the distance between carbon atoms) due to its impact on the potential energy of the system is one of the essential parameters that affects the electrical properties of the structure. In the following, we investigate the effects of loading, vacancy, and their combination. In the first step, at the edge of y=b, a force of $F=1\times10^{-9}$ N is applied to two atoms at $x_1$=3.69 nm, $x_2$=3.83 nm. By applying a tensile force to the pristine ribbon, the length of each change from L to L'. Then, three models of vacancies with the labels of A, B, and C are applied to the ribbon, and for simulating the vacancy, one carbon atom and elements attached to it are removed. Finally, the effect of mechanical loading in the presence of vacancy defects is investigated. It should be noted that the length of the bonds changes when the force is applied to the structure; namely, the new length of all elements is calculated.

By substituting the structural parameters that are obtained from Eq. (6) into Eq. (7), the stiffness matrix of the beam can be obtained [52]. The length of each element in the presence of a force, F, is calculated by applying the Hooke's law that is expressed as Eq. (9)

$$k_{11} = \begin{bmatrix} AS & 0 & 0 & 0 & 0 & 0 \\ 0 & a_z & 0 & 0 & 0 & b_z \\ 0 & 0 & a_y & 0 & -b_y & 0 \\ 0 & 0 & 0 & TS & 0 & 0 \\ 0 & 0 & -b_y & 0 & c_y & 0 \\ 0 & b_z & 0 & 0 & 0 & c_z \end{bmatrix}, \qquad (7)$$

$$k^e = \begin{bmatrix} k_{11} & -k_{11} \\ -k_{11} & k_{11} \end{bmatrix}, \qquad (8\text{-a})$$

$AS = \frac{EA}{L}$, $TS = \frac{GJ}{L}$, $a_z = \frac{12EI_z}{L^3}$, $b_z = \frac{6EI_z}{L^2}$, $c_z = \frac{4EI_z}{L}$, $d_z = \frac{2EI_z}{L}$, $a_y = \frac{12EI_y}{L^3}$, $b_y = \frac{6EI_y}{L^2}$,

$c_y = \frac{4EI_y}{L}$, $d_y = \frac{2EI_y}{L}$, (8-b)

$F = K^e \times \Delta L.$ (9)

In the stiffness matrix, EA, EI, and GJ represent the axial, bending, and torsional stiffness in the bonds, respectively.

### 2.2. Electrical modeling

In this subsection, we model the system using the TB approach [53, 54], and Harrison's scaling law. The electrical conductance of the AGNR is investigated in the cases of pure and defected [55]. Fig.1 (a) demonstrates a nanoribbon with a size of 7.53 nm×0.86 nm, including 288 carbon atoms, sandwiched between two electrodes. In general, the configuration is as follows, two semi-infinite electrodes with the armchair edge structure are connected to the ribbon at the edge of x=0 and x=a. The left and right electrodes and the ribbon are labeled as Source, Drain, and Device, respectively. Here, the device contains $N_A$=8 atoms in the y-direction and M=18 unit cells in the x-direction; namely, the total number of atoms of the device is 2M.$N_A$. Each unit cell contains $2N_A$ =16 carbon atoms (see the blue box in Fig.1).

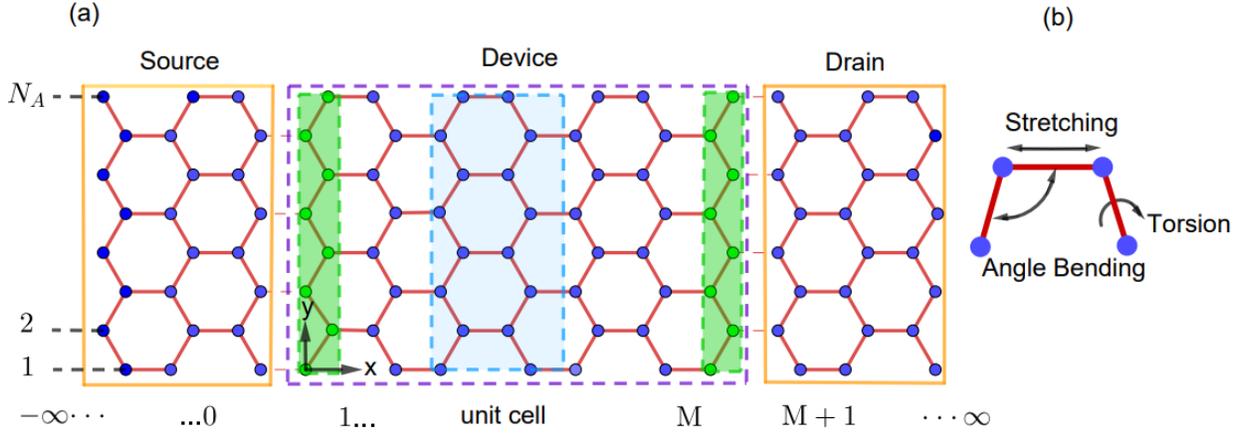

Fig. 1. (a) A schematic of the graphene-based armchair ribbon device, with a size of $N_A$=8 atoms and M=18 unit cells, connected to two semi-infinite electrodes. (b) Different interatomic interactions.

Here, the hopping energy for each bond length is calculated by applying the Harrison scaling law. The Hamiltonian matrices for the electrodes and device are introduced by the hopping energy t is set to -2.7 eV. Due to the dependence of the hopping energy on the c-c bond length, and after applying mechanical loading on the nanostructure, the bond lengths change; thus, the value of hopping energy in the device will no longer be equal to t=-2.7 eV, related to leads.

The Hamiltonian of the system is given by [53]:

$$H = \sum_j \varepsilon_j |j><j| + \sum_j (t_{j,j+1}|j><j+1| + t_{j+1,j}|j+1><j|). \tag{10}$$

where $\varepsilon_j$, $t_{j,j+1}$ are the on-site energy for atom $j$ and the hopping energy between atoms $j$ and $j+1$, respectively. When a force is applied to a system, due to the change in the length of bonds, new hopping energies can be calculated according to Harrison's scale law. The number of new hopping energies corresponds to the number of deformed elements. Therefore, they are given as [56]

$$t'^{(e)}_{j,j+1} = t_{j,j+1} \times (L^{(e)}/L'^{(e)}_{j,j+1})^2, \tag{11}$$

In Eq. (11), $t'^{(e)}_{j,j+1}$ and $L'^{(e)}_{j,j+1}$ are the new hopping energy and new length for each element, after deformation, respectively. Besides, after deformation, the new Hamiltonian matrix of the device is calculated. Now, we can start studying the transport properties of the system. For this, first, we need to get the source and drain surface Green's functions [53, 54]

$$g^s_{0,0} = [zI - H_{0,0} - H^\dagger_{-1,0}\tilde{T}]^{-1} \tag{12}$$

$$g^D_{M+1,M+1} = [zI - H_{M+1,M+1} - H_{M+1,M+2}T]^{-1} \tag{13}$$

Here, $H_{0,0(M+1,M+1)}$ represents the Hamiltonian for a unit cell of the source (drain) electrode, and $H_{-1,0(M+1,M+2)}$ is the coupling matrix between two unit cells in the source (drain) lead. Also, $z = E + i0^+$. The self-energies can be obtained as:

$$\Sigma_s(z) = H^\dagger_{0,1} g^s_{0,0} H_{0,1}, \tag{14}$$

$$\Sigma_D(z) = H_{M+1,M+1} g^D_{M+1,M+1} H^\dagger_{M,M+1}, \tag{15}$$

Eqs. (14) and (15) represent the self-energy matrix for coupling the source and drain electrodes to the device, respectively. The Green's function of the device can be calculated from

$$G(z) = [zI - H - \Sigma_s(z) - \Sigma_D(z)]^{-1} \tag{16}$$

Details of Eqs. (11) to (16) are in Refs. [53, 54, 57, 58].

## 3. Results and discussion

In this section, the effect of mechanical loading and vacancy defects on the structural properties of graphene nanoribbon and, finally, their effect on the electrical conductance of the system are investigated. As shown in Fig. 1(a), we assume that the width of the nanoribbon in the leads and the device is $N_A = 8$ atoms, and also, the device length is set to M = 18 armchair ribbon unit cells. We choose the on-site energy of carbon atoms equal to zero. Besides, the equilibrium Fermi energy of the leads is set to zero [59]

Many electronic properties of quasi-one-dimensional hexagonal nanomaterials are dependent on their edge structure and point defects. The electronic structure (red curves) and conductance (blue lines) of graphene-based armchair nanoribbon are illustrated in Fig. 2(a). The lowest-energy subbands indicate two metallic modes residing on the zigzag edges, and hence the system behaves as a metal (see red curves). The system has four channels for electron transmission; therefore, the maximum value of the transmission coefficient is 4. The allowable energies window is from -3t to 3t, namely, -8.1 to 8.1 eV. We can see that around the Fermi energy, the conductance and the density of states (Fig. 2 (b)) of the system are above zero, namely, confirmation of metallic behavior. Our numerical results are fully consistent with the results obtained by analytical solutions [60].

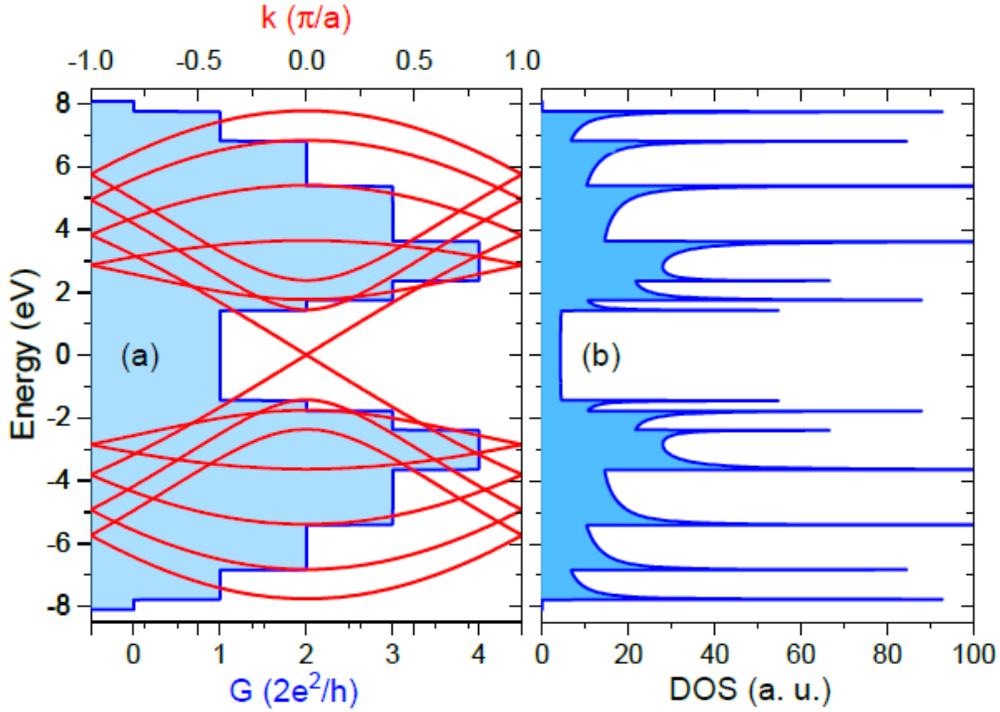

Fig. 2. (a) The variation of the conductance of the system as a function of the incident electron energy (blue lines), and the band structure (red curves). (b) The density of states of the system.

Fig. 3 illustrates the conductance spectra of structure, under three different forces, F, applied to two nodes of the system at positions $x_1$=3.692, $x_2$=3.834 nm, and $y_{1,2}$=b along the y- axis. Because of the applied force, the bond length between nearest-neighbor atoms $i$ and $j$ is changed as $L'^{(e)}_{j,j+1}$. We obtain this new bond length from the MM approach. From an experimental point of view, it is very difficult. The exact strain field, because of the deformation, is difficult to determine. Since the STM image has a limited resolution. Very recently, Li et al. [61] provided a simple ansatz for strain field approximation, which captures the leading order effects of the deformation. Fig. 3 (a) shows the conductance spectrum of the system under an applied force of F=1 nN (red curve). The value of vertical deflection due to this force is 0.262 nm, as shown in Fig. 3 (b). The dotted curve has been plotted as a reference, when the applied force is equal to zero. In the presence of this force, the electrical conductance of the system decreases due to the scattering of electrons from the deformed region. Besides, the conductance spectra vary from a stepped shape to an oscillating form. The reason for this is the carrier scattering when passing through the channel. According to panels (a), (c), and (e), the conductance of the structure decreases by increasing the magnitude of forces, namely, 1, 2, and 4 nN, but the nanostructure remains as a metal. We see this in figure 3 (a), (c), and, (e). Panels of (b), (d), and (f) in figure 3 display the spatial variation of the device, with vertical deflections of 0.262, 0.525, and 1.049 nm, respectively, under these three forces, which correspond to the left panels.

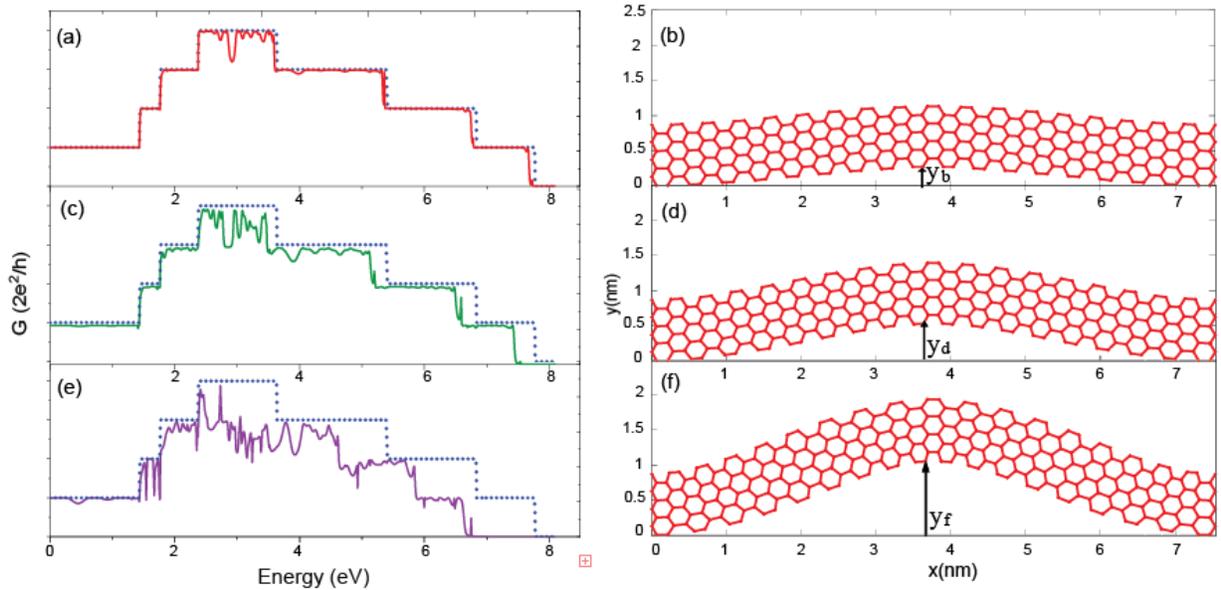

Fig. 3. The conductance spectra of the nanostructure under loading (a) F=1 nN, (c) F=2 nN, (e) F=4 nN. Right panels (b), (d), and (f) show the deformed nanostructure corresponding to the left panels. Here, $y_b$, $y_d$, and $y_f$ with the values of 0.262, 0.525, and 1.049 nm, respectively, show the maximum deflections of the structure after deformation. The dotted step-like curve has been plotted as a reference when the applied force is equal to zero.

In the following, the simultaneous effect of mechanical loading and vacancy defect on the conductance spectra of the structure is shown. To do this, we first introduce three different defective systems, namely, models A, B, and C, including three different configurations of the number of vacancy defects. Fig. 4 represents these three models. In model A, one vacancy has been placed at the center of the device. Also, models B and C contain 3 and 5 vacancies. The magnitude of the applied force is 2 nN. In right panels, $y_A$, $y_B$, and $y_C$ with the values of 0.525, 0.534, and 0.548 nm, respectively, are the maximum vertical deflections of the structure after applying the F. Here, under a fixed force, we observe that with increasing the number of vacancies, vertical deflections raise.

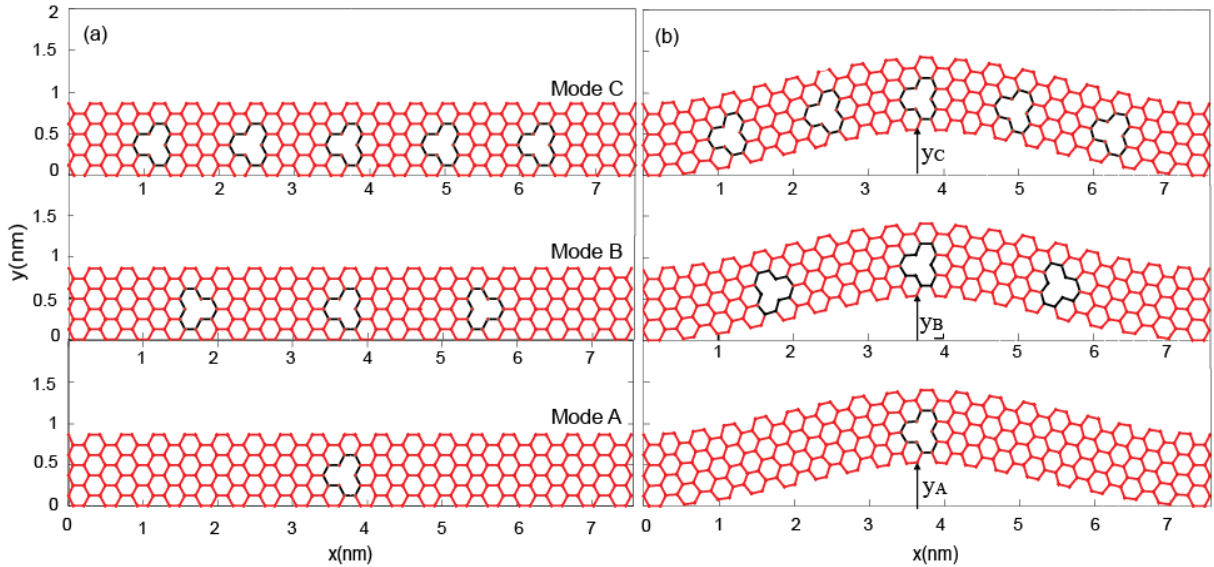

Fig. 4. (a) The schematic view of three different defective devices, namely, models A, B, and C, including one, three, and five vacancies, respectively. Here, we show the models without applying any force (left panels). Right panels, (b), show the models in the presence of a force of F=2 nN. $y_A$, $y_B$, and $y_C$ with values of 0.525, 0.534, and 0.548 nm, respectively, are the maximum vertical deflections of the structure after applying the F.

In Fig. 5, we show the conductance of nanoribbon with one vacancy defect, namely, Model A. It is clear that in the presence of a vacancy defect, the system behaves as a semiconductor with an energy gap of $E_g$=0.06 eV around the Fermi energy. Fig. 5 (a) depicts the conductance spectrum of the structure in the absence of any force. Besides, the panels of (b)-(d), related to the forces of 1, 2, and 4 nN, respectively, show the conductance spectra of the system. With increasing force, we can say that the size of the transport gap is almost fixed, but the conductance oscillations increases. This can be seen in Figure 5 (e).

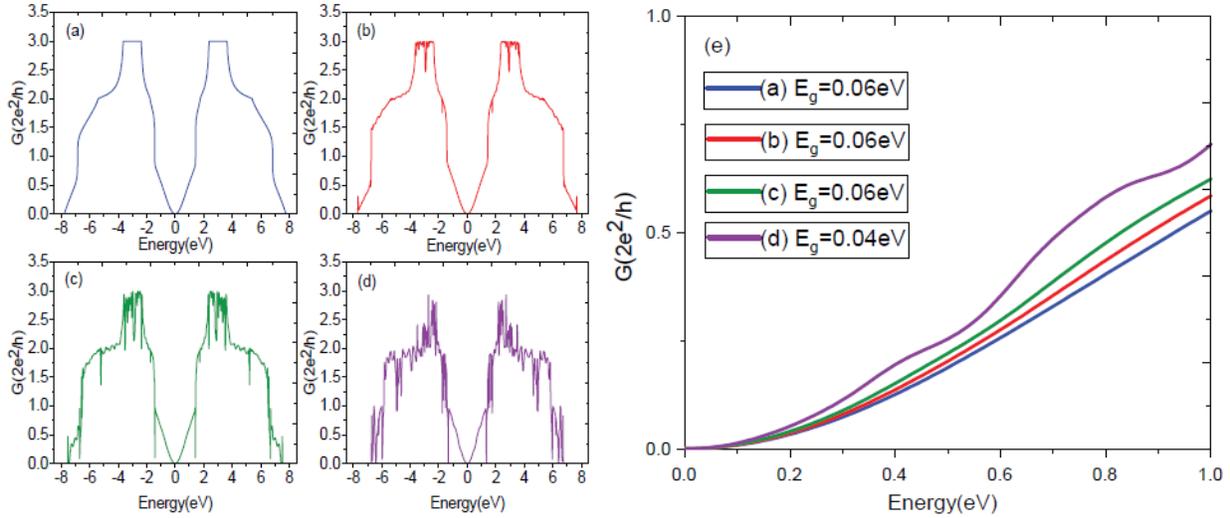

Fig. 5. The conductance of the system marked by model A as a function of electron energy. (a) F=0 N, (b) F=1 nN, (c) F=2 nN, (d) F=4 nN, (e) The high-resolution conductance of the system in the vicinity of the Fermi energy.

Fig. 6 presents the conductance spectra of the system, including three vacancies, namely, model B. The vacancies are at positions $x_1$=1.704, $x_2$=3.692, $x_3$=5.538 nm, and $y_{1,2,3}$=3.689 nm. According to this figure, the energy gap and the conductance decrease with increasing the applied force. Corresponding to the forces of 0, 1, 2, and 4 nN, the conductance gaps created in the system are equal to 0.56, 0.54, 0.51, and 0.46 eV, respectively. Beside, comparing figures 5 and 6, we find that as the number of vacancies increases, the size of the transport gap grows. Fig. 6 (a) depicts the conductance spectrum of the structure in the absence of any force. Besides, the panels of (b)-(d), related to the forces of 1, 2, and 4 nN, respectively, show the conductance spectra of the system. With increasing force, we can say that the size of the transport gap decreases, but the conductance oscillations increases. This can be clearly seen in Figure 6 (e). Generally, the width of the created band gap increases with increasing the number of vacancy defects and decreases with increasing the value of force, F. The position peaks of the graph represent the eigenvalues of the defective system under force.

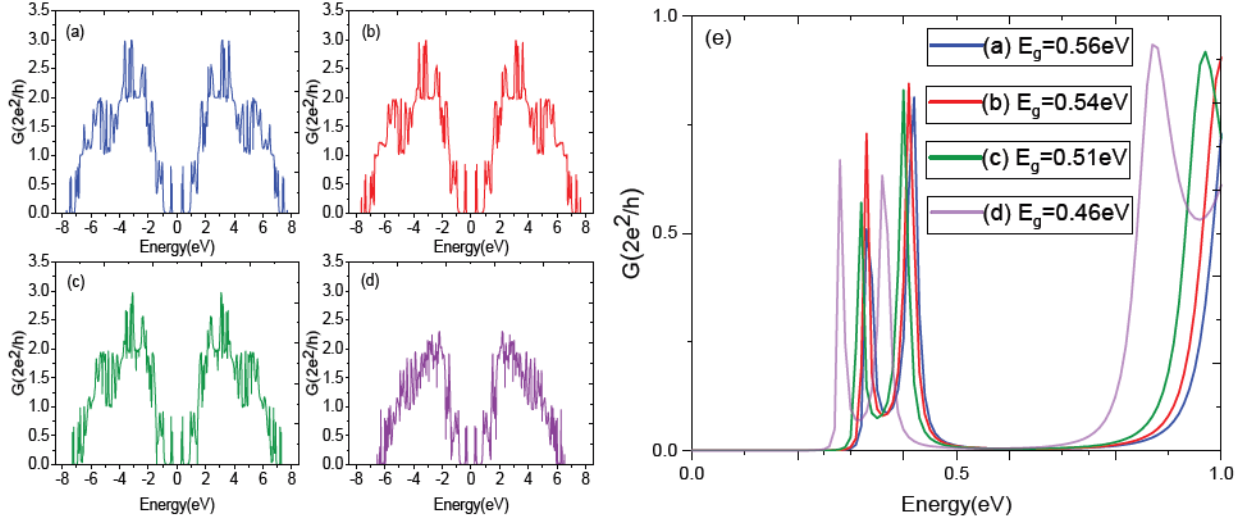

Fig. 6. The conductance of the system including three defects marked by model B. (a) F=0 N, (b) F=1 nN, (c) F=2 nN, (d) F=4 nN, (e) The high-resolution conductance of the system in the vicinity of the Fermi energy.

Finally, by increasing the number of vacancies to five, as labeled with model C, the transport gap of the system grows. As the number of point effects increases, the scattering of electrons increases when they pass through the device. The locations of these point defects are specified with coordinates $x_1$=1.136, $x_2$=2.414, $x_3$=3.692, $x_4$=4.97, $x_5$=6.248 nm, and $y_{1, 2,..., 5}$=3.689 nm. In Fig. 7, we illustrate the conductance spectra of the system, including five vacancies. According to this figure, the energy gap decreases with growing the magnitude of applied force. Namely, corresponding to the forces of 0, 1, 2, and 4 nN, the conductance gaps created in the system are equal to 1.50, 1.40, 1.34, and 1.20 eV, respectively. In addition, comparing figures 5, 6, and 7, we find that with increasing the number of point defects, the size of the transport gap rises. Generally, like fingerprints, these conductance spectra can be useful in detecting nano-displacements [62].

Briefly, in figure 8, we show the effect of external forces on the energy gap of the pure and the defective systems. Obviously, we see an increase in force reduces the energy gap for all three models A, B, and C.

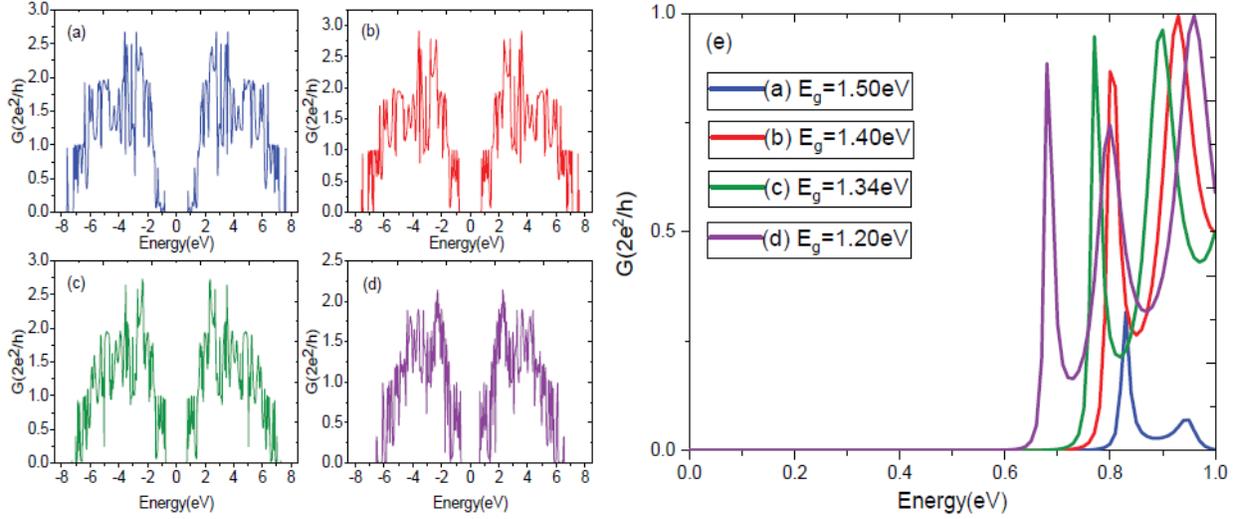

Fig. 7. The conductance spectra of the system with five defects marked by model C. (a) F=0 N, (b) F=1 nN, (c) F=2 nN, (d) F=4 nN, (e) The high-resolution conductance of the system in the vicinity of the Fermi energy.

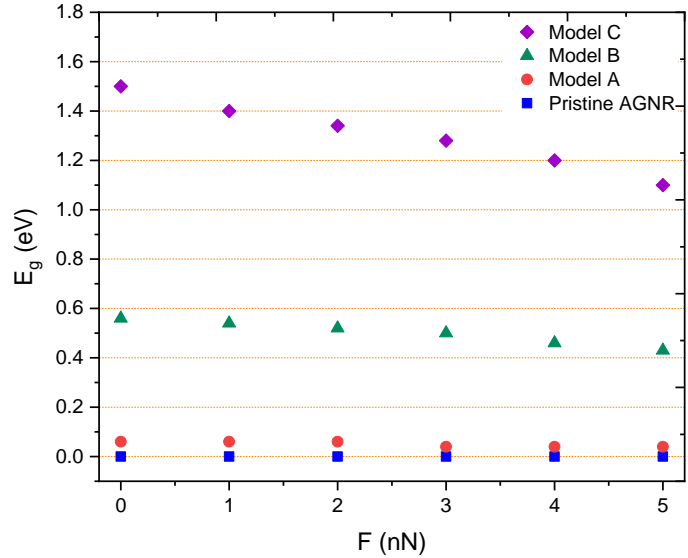

Fig. 8. The energy gaps of the system under four different conditions. The energy gap for model C is more significant than other models.

After examining the effect of force on the electronic transport properties of the system, it's time to look at the impact of the defect and **the direction of its extension as well as the effect of the ribbon width** on the electronic features of the system.

The changes of electronic behavior of structure in the presence of vacancy defects along ribbon direction have been reflected in Figs. (5), (6) and (7). According to these figures, the energy gap opened with a size of 0.06 eV in model A grows up to 1.5 eV in model C. Namely, the energy gap size is increasing with the number of defects. If the number of defects is much higher, the energy gap of the system increases to a certain limit of 1.56 eV for NA=8.

In Figure 9 (a), we show the effect of increasing the ribbon width on the energy gap of the system, when defects are present. The location of the defects does not change and their distance from the bottom edge of the ribbon remains unchanged. Thus, the increase in the width of the ribbon occurs through its upper edge. Blue, pink and green diagrams show the presence of one, three and five defects along the length of the system, the x- axis, respectively. The results show that the energy gap of the system increases as the number of defects grows. But these gaps decrease as the ribbon width increases.

Now, we want to examine the effect of the presence of one, three and five defects along the width of the ribbon, the y-axis, besides, the impact of the ribbon width on the energy gap of the system. Blue, pink and green diagrams show the presence of one, three and five defects along the y- axis, respectively. According to figure 9 (b), the energy gap size decreases to zero, as the width of the ribbon increases. When the width becomes very large, the ribbon also behaves similar to graphene and the energy gap reaches zero, which means that defects have no effect on large widths. In other words, as the width of the nanoribbon increases, the effect of the defect on the energy gap of the system decreases. Because, the number of electron channels of the system increases and the electrons can easily pass through the system. Therefore, when the width of nanoribbon becomes very large, the system turns into graphene and the energy gap decreases to zero, so that it will basically be a semimetal.

Furthermore, we examine the linear extension of these defects along the length of the ribbon. Carefully, in addition to investigating the effect of system width, we study the effect of the presence of one-, three- and five-line defects along the length of the ribbon. Our results are shown in Fig. 9 (c). Blue, pink and green curves show the presence of one-, three- and five- line defects. These diagrams accurately show **as the number of defects along the x- and or y-axes increase, the energy gap size of the system grows. But it decreases with increasing the width of the system.**

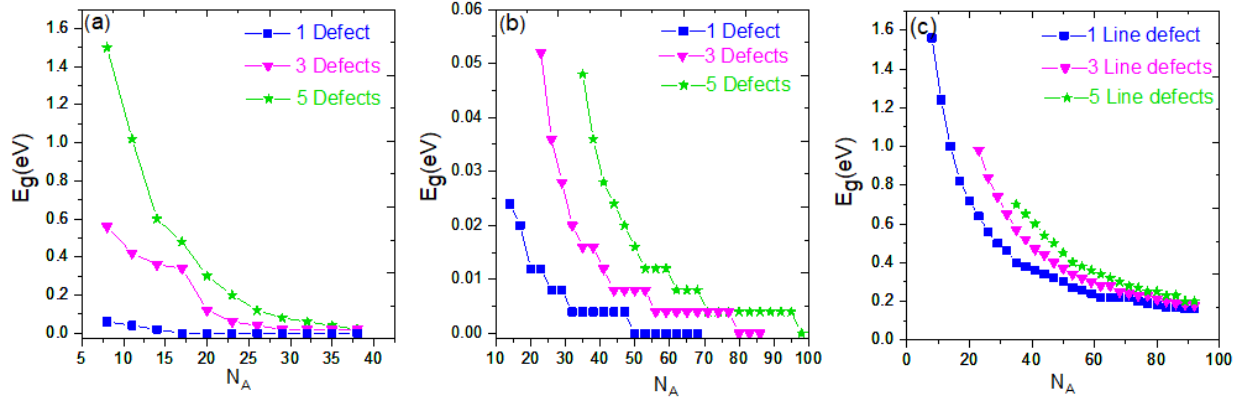

Fig. 9. The changes of energy gap as a function of the width of ribbon, (a) for models A, B and C. The defects are along the length of the ribbon, namely, the x-axis. (b) They are along the width of the ribbon, that is, the y-axis. (c) The bandgaps of middle panel in the case of line defects. Blue, pink and green curves show the presence of one-, three- and five-point and or line defects. The results show that in all cases the energy gap size decreases with increasing the width of the system.

## 4. Conclusions

In this paper, we proposed hybrid modeling of molecular mechanics and a tight-binding approach, which enables us to examine the effects of different parameters such as external local forces, constraints, and point defects on the electronic transport properties of nanostructures such as armchair graphene nanoribbons. Three models of vacancy defects applied to the structure, and the electrical conductance curves for the following cases obtained: a) applying local force to structure under constraint, b) applying vacancy, and c) applying vacancy and force to structure under constraint, simultaneously. Our results show that mechanical loading separately causes a reduction in the electrical conductance of the pure system, and the structure remains metal, but applying vacancy changes the metal structure to a semiconductor one. Also, by increasing the number of vacancies, the energy gap size of the strained system grows, and with increasing the magnitude of the force, the width of the transport gap decreases. In addition, in the absence of force, as the number of defects along the width and or length of the system increases, the energy gap size of the system grows. But it decreases with increasing the width of the system.

We can control the energy gap size and, consequently, the electrical properties of the system by tuning the vacancy defect concertation, its extension direction, and applying external local forces and constraints. The results of this research may be useful in the design of nanoelectromechanical systems and the detection of nano-displacements.

**Declaration of Competing Interest**

The authors declare that they have no competing interests.

Farhad Khoeini (khoeini@znu.ac.ir)